\documentclass[11pt]{article}

\pdfoutput=1

\usepackage[T1]{fontenc}
\usepackage[latin9]{inputenc}
\usepackage[a4paper]{geometry}
\usepackage[active]{srcltx}
\usepackage{amsmath}
\usepackage{amssymb}
\usepackage{esint}

\makeatletter

\usepackage{textcomp}

\pdfoutput=1 

\usepackage{jheppub}

\usepackage{etoolbox}
    
    \patchcmd{\maketitle}{\@fpheader}{}{}{}

\usepackage{amsfonts}

\usepackage{tikz}

\usepackage{bbm}

\title{\boldmath Hypergravity in five dimensions}

\author[a,b]{Oscar Fuentealba,}
\author[b]{Javier Matulich,}
\author[a]{and Ricardo Troncoso}

\affiliation[a]{Centro de Estudios Cient\'{i}ficos (CECs), Av. Arturo Prat 514, Valdivia,
Chile.}
\affiliation[b]{Universit\'e Libre de Bruxelles, and International Solvay Institutes, 
Campus Plaine --- CP 231, B-1050 Bruxelles, Belgium.}
\emailAdd{fuentealba@cecs.cl}
\emailAdd{jmatulic@ulb.ac.be}
\emailAdd{troncoso@cecs.cl}

\preprint{CECS-PHY-18/04}

\abstract{We show that a spin-$5/2$ field can be consistently coupled to 
gravitation without cosmological constant in five-dimensional spacetimes. 
The fermionic gauge ``hypersymmetry'' requires the presence of a finite 
number of additional fields, including a couple of $U(1)$ fields, a spinorial 
two-form, the dual of the graviton (of mixed $(2,1)$ Young symmetry) and 
a spin-$3$ field. The gravitational sector of the action is described by the purely quadratic
 Gauss-Bonnet term, so that the field equations for the metric are of 
second order. The local gauge symmetries of the full action principle close
 without the need of auxiliary fields. The field content corresponds to
the components of a connection for an extension of the 
``hyper-Poincar\'e'' algebra, which apart from the Poincar\'e and 
spin-$3/2$ generators, includes a generator of spin $2$ and a $U(1)$ 
central extension. It is also shown that this algebra admits an 
invariant trilinear form, which allows to formulate hypergravity as a gauge theory
described by a Chern-Simons action in five dimensions.}

\makeatother

\begin{document}
\maketitle \flushbottom

\section{Introduction}

Soon after the advent of supersymmetry, it was natural to wonder about
the possible superpartners of the graviton \cite{Salam:1974jj}. According
to the basic rules of supersymmetry, there are just two possibilities.
One of them corresponds to a massless spin-$3/2$ Rarita-Schwinger
field, which can be consistently coupled to gravitation. The full
theory is widely known as supergravity \cite{Freedman:1976xh}, \cite{Deser:1976eh}
and it has been extensively explored in different contexts (see e.g.,
\cite{VanNieuwenhuizen:1981ae}, \cite{Nilles:1983ge}, \cite{Haber:1984rc},
\cite{Martin:1997ns}, \cite{Freedman:2012zz}). One has barely heard
about the remaining possibility, which involves a massless field of
spin $5/2$, and there is a good reason for that: in four spacetime
dimensions the theory is inconsistent \cite{Aragone:1979hx}, \cite{Berends:1979kg}
(in agreement with the no-go theorems for interacting higher spin
theories, see e.g., \cite{Bekaert:2010hw}, \cite{Didenko:2014dwa},
\cite{Rahman:2015pzl}). The obstructions to achieve a consistent
coupling can be seen as follows. The action for a massless spin-$5/2$
field on flat space \cite{Schwinger:1970xc}, \cite{Fang:1978wz},
\cite{Berends:1979wu}, \cite{Curtright:1979uz}, \cite{Vasiliev:1980as}
possesses a fermionic gauge symmetry, that is not preserved once the
spin-$5/2$ field is minimally coupled to gravity \cite{Aragone:1979hx},
\cite{Berends:1979kg}. In few words, the variation of the fermionic
action under the fermionic gauge symmetry goes like the Riemann tensor,
while the variation of the Einstein-Hilbert action becomes proportional
to the Einstein tensor, leading no room for cancellation. Besides,
the consistency of the fermionic field equation implies the vanishing
of the Weyl tensor \cite{Aragone:1979hx}, \cite{Aragone:1980rk},
which is certainly too stringent as a condition to be imposed on four-dimensional
spacetimes. It has also been shown that neither the addition of cosmological
constant nor including non-minimal couplings help in order to circumvent
these obstructions \cite{Aragone:1979hx}, \cite{Aragone:1980rk},
\cite{Vasiliev:1986td}, \cite{Sorokin:2008tf} (see also e.g., \cite{Alkalaev:2002rq},
\cite{Ponomarev:2016lrm}, \cite{Skvortsov:2018jea}, \cite{Skvortsov:2018uru}).\footnote{Similar obstructions also occur for higher rank representations of
fields with spin $s\leq2$. In particular, in the case of a massless
spinorial two-form (see e.g., \cite{Townsend:1979yv}, \cite{Henneaux:2017xsb}),
fermionic gauge symmetries are also spoiled when the field is coupled
to gravitation \cite{Deser:1980rz}, \cite{Deser:1980fy}.} A well-known proposal to surmount these obstacles is given by Vasiliev's
theory \cite{Vasiliev:1990en}, \cite{Vasiliev:1999ba}, \cite{Vasiliev:2003ev}.
Its field equations describe a consistent coupling of a spin-$5/2$
field with gravitation in presence of cosmological constant, at the
expense of introducing an infinite tower of higher spin fields. Hitherto,
no standard action principle for Vasiliev's theory is known (see e.g.,
\cite{Boulanger:2011dd}), and since the spacetime metric is not invariant
under higher spin gauge symmetries, the theory should be ultimately
described by some sort of extension of Riemannian geometry.

In three spacetime dimensions the situation is radically different.
Indeed, the Weyl tensor identically vanishes in $D=3$, which allows
to express the Riemann tensor in terms of the Ricci tensor and the
Ricci scalar. Hence, the aforementioned obstructions to couple a spin-$5/2$
field to gravitation can be successfully bypassed.

\subsection{Hypergravity in three spacetime dimensions}

The consistent coupling of a spin-$5/2$ field to General Relativity
in $D=3$ was achieved by Aragone and Deser \cite{Aragone:1983sz},
who dubbed the theory as ``hypergravity''. It is the first and one
of the simplest theories describing a higher spin field coupled to
gravitation. The theory can be conveniently formulated in a local
frame, so that the action reads
\begin{equation}
I=\frac{k}{4\pi}\int2R^{a}e_{a}+i\bar{\psi_{a}}D\psi^{a}\,,\label{eq:Action3D}
\end{equation}
where $e^{a}$ stands for the dreibein, and $R^{a}=d\omega^{a}+\frac{1}{2}\epsilon^{abc}\omega_{b}\omega_{c}$
for the curvature two-form in terms of the dualized spin connection
$\omega^{a}=\frac{1}{2}\epsilon^{abc}\omega_{bc}$. The irreducible
spin-$5/2$ field $\psi^{a}=\psi_{\mu}^{a}dx^{\mu}$ is assumed to
be ``$\Gamma$-traceless'', i.e., $\Gamma^{a}\psi_{a}=0$, whose Lorentz
covariant derivative is given by $D\psi^{a}=d\psi^{a}+\frac{1}{2}\omega^{b}\Gamma_{b}\psi^{a}+\epsilon^{abc}\omega_{b}\psi_{c}$.\footnote{Wedge product between forms is assumed. We choose the orientation
so that the Levi-Civita symbol fulfills $\varepsilon_{012}=1,$ and
the Minkowski metric $\eta_{ab}$ follows the ``mostly plus'' convention.
In terms of the charge conjugation matrix $C$, the Majorana conjugate
reads $\bar{\psi}_{a\alpha}=\psi_{a}^{\beta}C_{\beta\alpha}$. An
imaginary unit ``$i$\textquotedblright{} in the product of real
Grassmann variables goes by hand with $\left(\theta_{1}\theta_{2}\right)^{*}=-\theta_{1}\theta_{2}$.
Our convention for (anti)symmetrization differs from that in \cite{Fuentealba:2015jma},
\cite{Fuentealba:2015wza}. Afterwards, round and square brackets
stand for symmetrization and antisymmetrization of the enclosed indices,
respectively, including a $1/2$ factor, i.e., $X^{(a|}Y^{b}Z^{|c)}=\frac{1}{2}(X^{a}Y^{b}Z^{c}+X^{c}Y^{b}Z^{a})$,
and $X^{[a|}Y^{b}Z^{|c]}=\frac{1}{2}(X^{a}Y^{b}Z^{c}-X^{c}Y^{b}Z^{a})$.
The three-dimensional Dirac matrices $\Gamma^{a}$ satisfy the Clifford
algebra $\{\Gamma_{a},\Gamma_{b}\}=2\eta_{ab}$.}

Following the ``1.5 formalism'' (see, e.g., \cite{VanNieuwenhuizen:1981ae})
it is simple to verify that the theory described by \eqref{eq:Action3D}
is invariant under
\begin{equation}
\delta\psi^{a}=D\epsilon^{a}\quad\text{and}\quad\delta e^{a}=\frac{3}{2}i\bar{\epsilon}_{b}\Gamma^{a}\psi^{b}\,,\label{eq:deltas3D-e-psi}
\end{equation}
where the spin-$3/2$ parameter $\epsilon^{a}=\epsilon^{a}(x^{\mu})$
is also $\Gamma$-traceless. In analogy with supergravity \cite{Deser:1976eh},
the spin connection was proposed to transform as \cite{Aragone:1983sz}
\begin{equation}
\delta\omega^{a}=3ie^{-1}\bar{\epsilon}^{b}\Gamma_{c}f_{b}^{\nu}\left(\frac{1}{2}e_{\nu}^{c}e^{a}-e_{\nu}^{a}e^{c}\right)\,,\label{eq:delta-omega-AD}
\end{equation}
with $f_{b}^{\nu}=\varepsilon^{\nu\rho\lambda}D_{\rho}\psi_{\lambda b}$.

In \cite{Fuentealba:2015jma}, it was shown that hypergravity can
be reformulated so that the field content corresponds to the components
of a gauge field that takes values in the ``hyper-Poincaré'' algebra
\begin{equation}
A=e^{a}P_{a}+\omega^{a}J_{a}+\psi_{a}^{\alpha}Q_{\alpha}^{a}\,,\label{eq:Connection3D}
\end{equation}
where the fermionic generators $Q_{\alpha}^{a}$ are $\Gamma$-traceless,
and its nonvanishing (anti)commutators read
\begin{gather}
\left[J_{a},J_{b}\right]=\varepsilon_{abc}J^{c}\quad,\quad\left[J_{a},P_{b}\right]=\varepsilon_{abc}P^{c}\,,\nonumber \\
\left[J_{a},Q_{\alpha b}\right]=\frac{1}{2}\left(\Gamma_{a}\right)_{\,\,\,\,\alpha}^{\beta}Q_{\beta b}+\varepsilon_{abc}Q_{\alpha}^{c}\,,\label{eq:hyperPoincare3D}\\
\left\{ Q_{\alpha}^{a},Q_{\beta}^{b}\right\} =-\frac{2}{3}\left(C\Gamma^{c}\right)_{\alpha\beta}P_{c}\eta^{ab}+\frac{5}{6}\varepsilon^{abc}C_{\alpha\beta}P_{c}+\frac{1}{3}(C\Gamma^{\left(a\right|})_{\alpha\beta}P^{\left|b\right)}\,.\nonumber 
\end{gather}
The hyper-Poincaré algebra \eqref{eq:hyperPoincare3D} admits an invariant
bilinear form, whose nonvanishing components are given by

\begin{equation}
\left\langle J_{a},P_{b}\right\rangle =\eta_{ab}\quad,\quad\left\langle Q_{\alpha}^{a},Q_{\beta}^{b}\right\rangle =\frac{2}{3}C_{\alpha\beta}\eta^{ab}-\frac{1}{3}\varepsilon^{abc}(C\Gamma_{c})_{\alpha\beta}\,,\label{eq:Brackets}
\end{equation}
and hence the hypergravity action \eqref{eq:Action3D} can be written
in terms of a Chern-Simons form
\begin{equation}
I=\frac{k}{4\pi}\int\left\langle AdA+\frac{2}{3}A^{3}\right\rangle ,\label{eq:CS-action3D}
\end{equation}
up to a boundary term. The local fermionic hypersymmetry is then spanned
by $\delta A=d\lambda+\left[A,\lambda\right]$, with $\lambda=\epsilon_{a}^{\alpha}Q_{\alpha}^{a}$,
so that
\begin{gather}
\delta\psi^{a}=D\epsilon^{a}\quad,\quad\delta\omega^{a}=0\quad,\quad\delta e^{a}=\frac{3}{2}i\bar{\epsilon}^{b}\Gamma^{a}\psi_{b}\,,\label{eq:deltas-CS-3D}
\end{gather}
which agree with \eqref{eq:deltas3D-e-psi}, \eqref{eq:delta-omega-AD}
only on-shell.

It is worth emphasizing that since the dynamical fields belong to
the components of the connection \eqref{eq:Connection3D}, instead
of a multiplet, the full set of transformation laws closes in the
hyper-Poincaré algebra without the need of auxiliary fields.

Having formulated hypergravity in terms of a Chern-Simons action certainly
helps in order to unveil the structure of the theory. Indeed, its
uniqueness has been recently established by virtue of BRST-cohomological
techniques \cite{Rahman:2019mra}. Besides, it was also possible to
analyze the asymptotic structure of the theory in \cite{Fuentealba:2015wza}.
The canonical generators of the asymptotic symmetries were shown to
span a hypersymmetric nonlinear extension of the BMS$_{3}$ algebra
(endowed with fermionic generators of conformal weight $5/2$), with
the same central extension as in the bosonic case \cite{Barnich:2006av}.
This hyper-BMS$_{3}$ algebra was also shown to admit unitary representations
\cite{Campoleoni:2015qrh}, \cite{Campoleoni:2016vsh}, \cite{Oblak:2016eij}.
Interestingly, as it occurs for $\mathcal{N}=1$ supergravity \cite{Barnich:2014cwa},
the anticommutator of the fermionic generators also allows to find
(an infinite number of) BPS-like bounds for the bosonic charges, which
in the hypersymmetric case turn out to be nonlinear. Circularly symmetric
solutions describing both cosmological spacetimes \cite{Ezawa:1992nk},
\cite{Cornalba:2002fi}, \cite{Cornalba:2003kd} and conical singularities
\cite{Deser:1983tn}, \cite{Deser:1984dr} fulfill all of the bounds
without saturating them (broken hypersymmetries), while configurations
with conical surpluses (angular excess) do not satisfy the bounds.
Hypersymmetry bounds saturate for configurations with unbroken hypersymmetries,
possessing globally well-defined Killing vector-spinors that fulfill
\begin{equation}
d\epsilon^{a}+\frac{1}{2}\omega^{b}\Gamma_{b}\epsilon^{a}+\epsilon^{abc}\omega_{b}\epsilon_{c}=0\,.
\end{equation}
In the case of fermions with antiperiodic boundary conditions, the
BPS-like configuration that is maximally hypersymmetric corresponds
to Minkowski spacetime; while for periodic boundary conditions it
is given by the null orbifold \cite{Horowitz:1990ap}, possessing
a single unbroken hypersymmetry.

It is also worth noting that, as in the case of supergravity \cite{Giacomini:2006dr},
\cite{Barnich:2014cwa}, \cite{Barnich:2015sca}, the hypergravity
theory can also be extended so as to include parity odd terms in the
action \cite{Fuentealba:2015jma}, \cite{Fuentealba:2015wza}, and
consequently, the asymptotic hyper-BMS$_{3}$ algebra becomes endowed
with an additional independent central extension along the Virasoro
subalgebra.

Hypergravity in 3D also admits a negative cosmological constant \cite{Chen:2013oxa},
\cite{Zinoviev:2014sza}, \cite{Henneaux:2015ywa} and it can be formulated
as a Chern-Simons theory for $osp(1|4)\oplus osp(1|4)$, at the expense
of including an additional spin-$4$ field and a Lorentz-like field
of mixed symmetry in tangent space. The local Lorentz algebra is then
enlarged from $sl(2,R)$ to $sp(4)$, so that the metric is not invariant
under spin-$4$ gauge symmetries. The asymptotic symmetries are given
by the direct sum of two copies of the $W_{(2,\frac{5}{2},4)}$ algebra,
from which nonlinear bounds for the bosonic charges can also be established
\cite{Henneaux:2015ywa}.\footnote{The analysis of $(M,N)$-extended hypergravity on $AdS_{3}$ and its
asymptotic structure has also been carried out in \cite{Henneaux:2015tar}.} In the limit of vanishing cosmological constant the spin-$4$ and
the Lorentz-like field can be consistently decoupled, so that hypergravity
can be suitably formulated in terms of Riemann-Cartan geometry. Furthermore,
as shown in \cite{Fuentealba:2015wza}, the hyper-BMS$_{3}$ algebra
also arises from a suitable truncation of $W_{(2,\frac{5}{2},4)}\oplus W_{(2,\frac{5}{2},4)}$
in the flat limit.

\subsection{Lifting to higher dimensions}

The formulation of hypergravity in 3D as a gauge theory of the hyper-Poincaré
algebra has been shown to be fruitful, and it is then natural to capture
some of its properties that could be lifted to higher dimensions.

In this sense, it is worth highlighting that the hyper-Poincaré algebra
actually exists for any $D>2$ dimensions \cite{Fuentealba:2015jma},
whose nonvanishing (anti)commutators read
\begin{eqnarray}
\left[J_{ab},J_{cd}\right] & = & J_{ad}\eta_{bc}-J_{bd}\eta_{ac}+J_{ca}\eta_{bd}-J_{cb}\eta_{ad}\nonumber \\
\left[J_{ab},P_{c}\right] & = & P_{a}\eta_{bc}-P_{b}\eta_{ac}\nonumber \\
\left[J_{ab},Q_{c}^{\alpha}\right] & = & -\frac{1}{2}(\Gamma_{ab})_{\,\,\,\,\beta}^{\alpha}Q_{c}^{\beta}+Q_{a}^{\alpha}\eta_{bc}-Q_{b}^{\alpha}\eta_{ac}\label{eq:hyperPoincareD}\\
\left[J_{ab},\bar{Q}_{\alpha c}\right] & = & \frac{1}{2}(\Gamma_{ab})_{\,\,\,\,\alpha}^{\beta}\bar{Q}_{\beta c}+\bar{Q}_{\alpha a}\eta_{bc}-\bar{Q}_{\alpha b}\eta_{ac}\nonumber 
\end{eqnarray}
\begin{equation}
\left\{ Q^{\alpha a},\bar{Q}_{\beta}^{b}\right\} =\frac{3\left(D-2\right)}{D^{2}}i\left[\left(D+1\right)(\Gamma^{c})_{\,\,\,\,\beta}^{\alpha}P_{c}\eta^{ab}-\frac{D+2}{D-2}(\Gamma^{abc})_{\,\,\,\,\beta}^{\alpha}P_{c}-2(\Gamma^{\left(a\right|})_{\,\,\,\,\beta}^{\alpha}P^{\left|b\right)}\right]\label{eq:QQhyperPoincareD}
\end{equation}
where $\bar{Q}_{a}=Q_{a}^{\dagger}\Gamma^{0}$ generically stands
for the Dirac conjugate.\footnote{As in $D=3$, the hyper-Poincaré algebra can also be readily written
in terms of Majorana spinors when they exist. In two spacetime dimensions
the algebra is still consistent, but nevertheless, the subset spanned
by translations and fermionic generators becomes an Abelian ideal.}

Thus, the very existence of the algebra \eqref{eq:hyperPoincareD},
\eqref{eq:QQhyperPoincareD} suggests the possibility that hypergravity
in higher dimensions might be formulated as a gauge theory of some
suitable extension of the hyper-Poincaré group. Indeed, if such extension
admitted an invariant tensor of rank $n$, a non-Abelian Chern-Simons
action in $D=2n-1$ dimensions might be a good candidate to explore.

Another interesting lesson that can be extracted from the three-dimensional
case without cosmological constant is that, according to the action
principle in \eqref{eq:Action3D}, the spin-$5/2$ field couples to
the geometry exclusively through the spin connection (the dreibein
is not involved in the Lorentz covariant derivative). Hence, in our
formulation, the fermionic field does not contribute to the stress-energy
tensor, but it is noticed by the geometry due to the torsion $T^{a}=\frac{3}{4}i\bar{\psi}_{b}\Gamma^{a}\psi^{b}$.
Such type of couplings are certainly excluded in $D=4$ \cite{Aragone:1980rk},
but nonetheless, there is some evidence suggesting that this kind
of couplings could be realized in higher odd dimensions. Indeed, a
class of supergravity theories featuring fermions that are non-minimally
coupled to the curvature, but not to the vielbein, is known to exist
for any odd dimensions \cite{Banados:1996hi}, from which one can
also extract a helpful moral.

For the sake of simplicity, let us consider the case of (nonstandard)
supergravity in five dimensions, whose action reads
\begin{equation}
I=I_{GB}-3I_{b}+I_{\psi}\,,\label{eq:Action1-1}
\end{equation}
where
\begin{eqnarray}
I_{GB} & = & \frac{1}{2}\int\varepsilon_{abcdf}R^{ab}R^{cd}e^{f}\,,\label{eq:Gauss-Bonnet}\\
I_{b} & = & \int R^{ab}R_{ab}b\,,\label{eq:I_b}\\
I_{\psi} & = & 3\int\bar{\psi}R^{bc}\Gamma_{bc}D\psi+h.c\,.\label{eq:I_fermion}
\end{eqnarray}
Here, the gravitational sector is described by the pure Gauss-Bonnet
action $I_{GB}$ in \eqref{eq:Gauss-Bonnet}. Since $I_{GB}$ corresponds
to a particular case of Lovelock theory, devoid of the Einstein-Hilbert
and cosmological terms, the field equations for the metric are of
second order.

Invariance under local supersymmetry requires the presence of an additional
bosonic term $I_{b}$ \eqref{eq:I_b}, where $b=b_{\mu}dx^{\mu}$
stands for an Abelian 1-form.

The dynamics of the spin-$3/2$ field $\psi=\psi_{\mu}dx^{\mu}$,
whose Lorentz covariant derivative is given by $D\psi=d\psi+\frac{1}{4}\omega^{ab}\Gamma_{ab}\psi$,
is described by the fermionic term $I_{\psi}$ in \eqref{eq:I_fermion}.
Note that the fermionic field non-minimally couples to the spacetime
geometry through the curvature two-form $R^{ab}$ (instead of the
vielbein).

The action \eqref{eq:Action1-1} is invariant under the following
local supersymmetry transformations
\begin{gather}
\delta\psi^{a}=D\epsilon^{a}\qquad,\qquad\delta\omega^{ab}=0\,,\label{eq:trans-sugra5D}\\
\delta e^{a}=3i\bar{\epsilon}\Gamma^{a}\psi+h.c.\qquad,\qquad\delta b=\bar{\text{\ensuremath{\epsilon}}}\psi+h.c.
\end{gather}

It is worth noting that since $I_{GB}$ in \eqref{eq:Gauss-Bonnet}
is linear in the vielbein instead of the curvature, the local Lorentz
symmetry is extended to the local Poincaré group in five dimensions.
The full set of local symmetries also includes the gauge transformation
associated to the bosonic Abelian field $b$, and hence, their algebra
corresponds to super-Poincaré with a $U(1)$ central extension, so
that the anticommutator of the fermionic generators reads
\begin{equation}
\{Q^{\alpha},\bar{Q}_{\beta}\}=3i(\Gamma^{a})_{\,\,\,\beta}^{\alpha}P_{a}+\delta_{\beta}^{\alpha}K\,.
\end{equation}
The presence of the central extension plays a relevant role, since
in this case the super-Poincaré algebra admits an invariant (anti-)symmetric
form of rank three, whose nonvanishing components are given by
\begin{eqnarray}
\left\langle J_{ab},J_{cd},P_{f}\right\rangle  & = & \frac{2}{3}\varepsilon_{abcdf}\,,\\
\left\langle J_{ab},J_{cd},K\right\rangle  & = & -4\eta_{a\left[c\right.}\eta_{\left.d\right]b}\,,\\
\left\langle Q^{\alpha},J_{ab},\bar{Q}_{\beta}\right\rangle  & = & -2(\Gamma_{ab})_{\,\,\,\beta}^{\alpha}\,,\nonumber 
\end{eqnarray}
which is crucial in order to formulate this class of supergravity
as a gauge theory. Indeed, the field content corresponds to the components
of a connection for the centrally extended super-Poincaré algebra
\begin{equation}
A=e^{a}P_{a}+\frac{1}{2}\omega^{ab}J_{ab}+bK+\bar{\psi}Q-\bar{Q}\psi\,,
\end{equation}
so that the local (super)symmetries are obtained from a gauge transformation
$\delta A=d\lambda+[A,\lambda]$, where $\lambda$ takes values on
the super-Poincaré algebra with a $U(1)$ central extension. Thus,
up to a boundary term, the supergravity action \eqref{eq:Action1-1}
can be written as a Chern-Simons form
\begin{equation}
I=\int\left\langle AF^{2}-\frac{1}{2}A^{3}F+\frac{1}{10}A^{5}\right\rangle \,,\label{eq:CS-5D}
\end{equation}
where $F=dA+A^{2}$.

In the next section we construct a hypergravity theory in five dimensions
that shares some of these features.

\section{Hypergravity in five dimensions}

Following the moral outlined in the introduction, for a fermionic
field $\psi^{a}=\psi_{\mu}^{a}dx^{\mu}$ that fulfills $\Gamma^{a}\psi_{a}=0$,
we look for a kinetic term whose coupling to the geometry does not
involve the vielbein, being of the form 
\begin{equation}
\bar{\psi}_{a}X^{ab}D\psi_{b}+h.c.
\end{equation}
Local Lorentz invariance then implies that $X^{ab}$ must be necessarily
proportional to the curvature two-form $R^{cd}$. It is then simple
to verify that there are just five possibilities, so that $X^{ab}$
is given by a combination of the following terms:
\begin{gather}
R^{ab}\,\,,\,\,\eta^{ab}R^{cd}\Gamma_{cd}\label{eq:2poss}\\
R_{\,\,\,c}^{a}\Gamma^{cb}\,\,,\,\,\Gamma_{\,\,\,c}^{a}R^{cb},\,\,\epsilon^{abcde}R_{cd}\Gamma_{e}\label{eq:3poss}
\end{gather}
Nonetheless, since the fermionic field is $\Gamma$-traceless, the
last three possibilities in \eqref{eq:3poss} become redundant, and
the relevant ones then reduce to the remaining two in \eqref{eq:2poss}.
Hence, $X^{ab}$ must be of the form
\begin{equation}
X_{\,\,\,b}^{a}=\delta_{\,\,\,b}^{a}R^{cd}\Gamma_{cd}+\alpha R_{\,\,\,b}^{a}\,.\label{eq:Xab}
\end{equation}
We also assume that the local fermionic gauge symmetry (hypersymmetry)
is spanned by a spin-$3/2$ parameter $\epsilon^{a}$ subject to the
$\Gamma$-traceless condition ($\Gamma^{a}\epsilon_{a}=0$), so that
the transformation law of the fermionic field and the spin connection
agree with those in the three-dimensional case, i.e.,
\begin{equation}
\delta\psi^{a}=D\epsilon^{a}\qquad,\qquad\delta\omega^{ab}=0\,.\label{eq:trans-psi-omega}
\end{equation}

As explained below, it is useful to fix the arbitrary constant in
$X_{\,\,\,b}^{a}$ according to $\alpha=-4$, since it minimizes the
number of additional bosonic fields, and in particular, it avoids
the introduction of extra fields with mixed symmetry on tangent space.

\subsection{Action principle and local (hyper)symmetries}

We are then led to propose the following action principle
\begin{equation}
I=I_{GB}+\frac{9}{5}I_{b}+I_{e^{ab}}+I_{\psi^{a}}\,,\label{eq:Action1}
\end{equation}
where $I_{GB}$ stands for the pure Gauss-Bonnet action in \eqref{eq:Gauss-Bonnet},
$I_{b}$ is given by \eqref{eq:I_b}, and the remaining terms read
\begin{eqnarray}
I_{e^{ab}} & = & -24\int R_{\,\,\,c}^{a}R^{cb}e_{ab}\,,\\
I_{\psi^{a}} & = & 3\int\bar{\psi}_{a}\left(\delta_{b}^{a}R^{cd}\Gamma_{cd}-4R_{\,\,\,b}^{a}\right)D\psi^{b}+h.c.\label{eq:Fermionic-action}
\end{eqnarray}
Apart from the vielbein $e^{a}$, the spin connection $\omega^{ab}$
and the fermionic one-form field $\psi^{a}$, the theory includes
additional bosonic fields described by an Abelian field $b=b_{\mu}dx^{\mu}$,
and a one-form given by $e^{ab}=e_{\mu}^{ab}dx^{\mu}$, being symmetric
and traceless in tangent space ($e^{ab}=e^{ba}$, $\eta_{ab}e^{ab}=0$).

Looking at the irreducible components, the fermionic one-form contains
a spin-$5/2$ field $\psi_{(\mu\nu)}$ and a spinorial two-form $\psi_{[\mu\nu]}$,
which are symmetric and antisymmetric under $\mu\leftrightarrow\nu$,
respectively.\footnote{Note that in four-dimensional spacetimes the coupling of each of these
fermionic fields with gravitation becomes inconsistent \cite{Aragone:1979hx},
\cite{Deser:1980rz}. It is also worth pointing out that the fermionic
action \eqref{eq:Fermionic-action} appears to be devoid of an analog
of the local ``shift symmetry'' \cite{Vasiliev:1986td}, \cite{Sorokin:2008tf},
\cite{Rahman:2017cxk} that allows to consistently gauge away the
spinorial two-form from the analysis.} Analogously, the one-form $e^{ab}$ decomposes as
\begin{equation}
e_{\mu}^{ab}:\,\,\{e_{(\mu\nu\lambda)}\,\,,\,\,e_{\,\,\,\nu\mu}^{\nu}\,\,,\,\,e_{\mu\nu,\lambda}\}\,,
\end{equation}
so that its trace $e_{\,\,\,\nu\mu}^{\nu}$ corresponds to a spin-$1$
field, while the fully symmetric part $e_{(\mu\nu\lambda)}$ describes
the highest spin field ($s=3$). The remaining field $e_{\mu\nu,\lambda}$
is of mixed (2,1) Young symmetry, remarkably coinciding with the dual
of the graviton in five dimensions \cite{Curtright:1980yk} (see also
\cite{Henneaux:2019zod}). It is worth highlighting that in our context,
the field $e_{\mu\nu,\lambda}$ is strictly required by hypersymmetry,
instead of duality at the linearized level.

We should also emphasize one of the advantages of keeping all of the
irreducible components of the fields aforementioned. Indeed, if one
wanted to project them out, one might follow the nowadays standard
procedure of enlarging the Lorentz group by introducing additional
fields of mixed symmetry in tangent space (see e.g., \cite{Vasiliev:1980as},
\cite{Vasiliev:1986td}, \cite{Lopatin:1987hz}, \cite{Campoleoni:2010zq},
\cite{Alkalaev:2010af}, \cite{Rahman:2017cxk}). However, when the
spacetime geometry becomes dynamical, the price to pay would be that
the metric transforms in a non-trivial way under higher spin transformations,
implying that Riemannian geometry should be extended in some suitable
way, which is hitherto unknown. Hence, keeping all of the irreducible
components aforementioned, allows us to formulate a consistent theory
of hypergravity in five dimensions in a conservative way, described
by a local field theory with a finite number of fields (of spin up
to three) on standard Riemann-Cartan geometry.

The action is invariant under the following local hypersymmetry transformations
\begin{gather}
\delta\psi^{a}=D\epsilon^{a}\quad,\quad\delta\omega^{ab}=0\,.\label{eq:trans-psi-omega-1}\\
\delta e^{a}=3i\bar{\epsilon}^{b}\Gamma^{a}\psi_{b}+h.c.\label{eq:trans-psi-omega-2}\\
\delta b=\bar{\epsilon}^{a}\psi_{a}+h.c.\\
\delta e^{ab}=\bar{\epsilon}^{(a}\psi^{b)}-\frac{1}{5}\eta^{ab}\bar{\epsilon}_{c}\psi^{c}+h.c.
\end{gather}
Note that the transformations in \eqref{eq:trans-psi-omega-1}, \eqref{eq:trans-psi-omega-2}
agree those in the three-dimensional case (extended for Dirac fermions).

Apart from the manifest local Lorentz invariance, the action \eqref{eq:Action1}
is also invariant under local translations, and $U(1)$ gauge transformations
with parameters $\lambda^{a}$ and $\lambda$, respectively. An additional
local bosonic ``spin 1-2-3'' symmetry is spanned by a parameter
$\lambda^{ab}$, that is symmetric and traceless. The transformation
law of the fields under these symmetries are given by
\begin{gather}
\delta e^{a}=D\lambda^{a}\quad,\quad\delta b=d\lambda\quad,\quad\delta e^{ab}=D\lambda^{ab}\,,\\
\delta\omega^{ab}=0\quad,\quad\delta\psi^{a}=0\,.
\end{gather}

The whole set of local gauge symmetries of the hypergravity action
\eqref{eq:Action1} turns out to close for an extension of the hyper-Poincaré
algebra in \eqref{eq:hyperPoincareD}, \eqref{eq:QQhyperPoincareD}
in five dimensions, endowed with an additional $U(1)$ generator $K$
and a symmetric and traceless spin$-2$ generator $P_{ab}$, whose
nonvanishing (anti)commutators manifestly read
\begin{eqnarray}
\left[J_{ab},J_{cd}\right] & = & J_{ad}\eta_{bc}-J_{bd}\eta_{ac}+J_{ca}\eta_{bd}-J_{cb}\eta_{ad}\,,\nonumber \\
\left[J_{ab},P_{c}\right] & = & P_{a}\eta_{bc}-P_{b}\eta_{ac}\,,\nonumber \\
\left[J_{ab},P_{cd}\right] & = & P_{ad}\eta_{bc}-P_{bd}\eta_{ac}+P_{ca}\eta_{bd}-P_{cb}\eta_{ad}\,,\nonumber \\
\left[J_{ab},Q_{c}^{\alpha}\right] & = & -\frac{1}{2}\left(\Gamma_{ab}\right)_{\,\,\,\beta}^{\alpha}Q_{c}^{\beta}+Q_{a}^{\alpha}\eta_{bc}-Q_{b}^{\alpha}\eta_{ac}\,,\label{eq:hyper-Poincare-extended}\\
\left[J_{ab},\bar{Q}_{\alpha c}\right] & = & \frac{1}{2}\left(\Gamma_{ab}\right)_{\,\,\,\alpha}^{\beta}\bar{Q}_{\beta c}+\bar{Q}_{\alpha a}\eta_{bc}-\bar{Q}_{\alpha b}\eta_{ac}\,,\nonumber \\
\left\{ Q^{\alpha a},\bar{Q}_{\beta}^{b}\right\}  & = & \frac{54}{25}i(\Gamma^{c})_{\,\,\,\beta}^{\alpha}P_{c}\eta^{ab}-\frac{21}{25}i(\Gamma^{abc})_{\,\,\,\beta}^{\alpha}P_{c}-\frac{18}{25}i(\Gamma^{\left(a\right|})_{\,\,\,\beta}^{\alpha}P^{\left|b\right)}\nonumber \\
 &  & +\frac{1}{5}\left(4\eta^{ab}\delta_{\beta}^{\alpha}-(\Gamma^{ab})_{\,\,\,\beta}^{\alpha}\right)K+\frac{1}{10}\left(3\delta_{\beta}^{\alpha}P^{ab}+2(\Gamma_{c}^{\,\,\,\left[a\right|})_{\,\,\,\beta}^{\alpha}P^{\left|b\right]c}\right)\,.\nonumber 
\end{eqnarray}
\newpage

\subsection{Hypergravity as a gauge theory for the extended hyper-Poincaré algebra}

Interestingly, the local gauge symmetries of the hypergravity action
close according to the extended hyper-Poincaré algebra in \eqref{eq:hyper-Poincare-extended}
without the need of auxiliary fields. Indeed, this is a consequence
of the fact that the field content precisely corresponds to the components
of a single connection of the extended hyper-Poincaré algebra
\begin{equation}
A=e^{a}P_{a}+\frac{1}{2}\omega^{ab}J_{ab}+bK+\frac{1}{2}e^{ab}P_{ab}+\bar{\psi}^{a}Q_{a}-\bar{Q}_{a}\psi^{a}\,,\label{Connection_hygra}
\end{equation}
and hence, the local gauge symmetries are recovered from $\delta A=d\eta+[A,\eta]$,
where $\eta$ stands for an extended hyper-Poincaré-algebra-valued
zero-form
\begin{equation}
\eta=\lambda^{a}P_{a}+\frac{1}{2}\sigma^{ab}J_{ab}+\lambda K+\frac{1}{2}\lambda^{ab}P_{ab}+\bar{\epsilon}^{a}Q_{a}-\bar{Q}_{a}\epsilon^{a}\,.
\end{equation}

Besides, the additional bosonic generators $K$ and $P_{ab}$ extend
the hyper-Poincaré algebra so that it admits an invariant (anti-)symmetric
form of rank three. Their nonvanishing components are given by
\begin{eqnarray}
\left\langle J_{ab},J_{cd},P_{f}\right\rangle  & = & \frac{2}{3}\varepsilon_{abcdf}\,,\nonumber \\
\left\langle J_{ab},J_{cd},K\right\rangle  & = & \frac{12}{5}\eta_{a\left[c\right.}\eta_{\left.d\right]b}\,,\label{eq:bracket-5D-hyper}\\
\left\langle J_{ab},J_{cd},P_{ef}\right\rangle  & = & -64\left(\eta_{e\left[a\right.}\eta_{\left.b\right]\left[c\right.}\eta_{\left.d\right]f}-{\displaystyle \frac{1}{5}}\eta_{c\left[a\right.}\eta_{\left.b\right]d}\eta_{ef}\right)\,,\nonumber \\
\left\langle Q_{c}^{\alpha},J_{ab},\bar{Q}_{\beta d}\right\rangle  & = & 2[4\mathbb{P}_{c\left[a\right.}\mathbb{P}_{\left.b\right]d}-\mathbb{P}_{ce}\Gamma_{ab}\mathbb{P}_{\,\,\,d}^{e}]_{\,\,\,\beta}^{\alpha}\,,\nonumber 
\end{eqnarray}
where the $\Gamma$-traceless projector $(\mathbb{P}_{ab})_{\,\,\,\beta}^{\alpha}=\eta_{ab}\delta_{\,\,\,\beta}^{\alpha}-\frac{1}{5}\left(\Gamma_{a}\Gamma_{b}\right)_{\,\text{\,\,\ensuremath{\beta}}}^{\alpha}$,
ensures $\Gamma$-traceless of the fermionic entries of the bracket.

Therefore, the hypergravity theory \eqref{eq:Action1} can be formulated
as a gauge theory with standard fiber bundle structure. Indeed, by
virtue of \eqref{Connection_hygra} and \eqref{eq:bracket-5D-hyper},
the action \eqref{eq:Action1} can be written in terms of a Chern-Simons
form in five dimensions
\begin{equation}
I=\int\left\langle AF^{2}-\frac{1}{2}A^{3}F+\frac{1}{10}A^{5}\right\rangle \,,\label{eq:CS-5D-1}
\end{equation}
up to a boundary term. Note that the field equations can then be compactly
expressed in a manifestly covariant way under the extended hyper-Poincaré
algebra, spanned by the set $G_{I}=\left\{ J_{ab},P_{a},K,P_{ab},Q_{a},\bar{Q}_{a}\right\} $,
according to
\begin{equation}
\left\langle F^{2}G_{I}\right\rangle =0\,.\label{eq:FIeld-equations}
\end{equation}
Here, $F=dA+A^{2}$ stands for the field strength, whose components
are given by
\begin{equation}
F=\tilde{T}^{a}P_{a}+\frac{1}{2}R^{ab}J_{ab}+\tilde{T}K+\frac{1}{2}\tilde{T}^{ab}P_{ab}+D\bar{\psi}^{a}Q_{a}-\bar{Q}_{a}D\psi^{a}\,,\label{eq:F-extended}
\end{equation}
where
\begin{eqnarray}
\tilde{T}^{a} & = & T^{a}-3i\bar{\psi}_{b}\Gamma^{a}\psi^{b}\,,\\
\tilde{T} & = & db-\bar{\psi}_{a}\psi^{a}\,,\\
\tilde{T}^{ab} & = & De^{ab}-\bar{\psi}^{(a}\psi^{b)}+\frac{1}{5}\eta^{ab}\bar{\psi}_{c}\psi^{c}\,,
\end{eqnarray}
and $T^{a}=De^{a}$ is the torsion.

\section{Ending remarks}

We have proposed an action principle for hypergravity in five-dimensional
spacetimes, that can be formulated as a field theory with standard
fiber bundle structure. Noteworthy, the theory contains a finite number
of higher spin fields (up to just $s=3$), that can be seen as the
components of a connection for the extended hyper-Poincaré group.
Thus, as the fields are not arranged within the components of an irreducible
multiplet, the possible inconsistencies associated to the Haag-\L opusza\'{n}ski--Sohnius
theorem \cite{Haag:1974qh} can be successfully circumvented. In particular,
bosonic and fermionic degrees of freedom are then not restricted to
match.

The theory possesses a simple geometric structure, but as it is the
case of generic non-Abelian Chern-Simons theories in five dimensions,
the analysis of its dynamics is not so straightforward \cite{Banados:1995mq},
\cite{Banados:1996yj}, \cite{Chandia:1998uf}, \cite{Miskovic:2005di},
\cite{Miskovic:2006ei}, since the constraint structure changes along
phase space. A simple way of visualizing this is the following. The
field equations \eqref{eq:FIeld-equations} are trivially solved by
Minkowski spacetime in vacuum, because in that case the field strength
$F$ in \eqref{eq:F-extended} vanishes. Nevertheless, since the field
equations are purely quadratic in $F$, linear perturbations around
this (maximally hypersymmetric) configuration possess a ``linearization
instability'', implying that the analysis necessarily requires to
go to higher order. As pointed out in \cite{Hassaine:2003vq}, \cite{Hassaine:2004pp},
this feature appears to be welcome, since the theory naturally tends
to explore different vacua. Note that solutions in vacuum, without
torsion, fulfill the analogue of the Einstein equation
\begin{equation}
\varepsilon_{abcdf}R^{ab}R^{cd}=0\,,\label{eq:R2-Einstein}
\end{equation}
with an additional condition on the geometry that comes from the variation
of the action \eqref{eq:Action1} with respect to the bosonic fields
$b$ and $e_{ab}$, given by
\begin{equation}
R_{\,\,\,c}^{a}R^{cb}=0\,.\label{eq:Constraint-geometry}
\end{equation}
Equations \eqref{eq:R2-Einstein} and \eqref{eq:Constraint-geometry}
also emerge from the consistency of the fermionic field equations.
It is then worth highlighting that the latter condition is not as
stringent as requiring the vanishing of the Weyl tensor. Indeed, it
is simple to verify that the spherically symmetric solution of \eqref{eq:R2-Einstein}
(see \cite{Banados:1993ur}, \cite{Crisostomo:2000bb}) automatically
fulfills \eqref{eq:Constraint-geometry}, without imposing any additional
condition on the geometry.

The field equations \eqref{eq:R2-Einstein} and \eqref{eq:Constraint-geometry}
can also be seen to admit braneworld-like solutions in vacuum, whose
metric is given by \cite{Hassaine:2003vq}
\begin{eqnarray}
ds^{2} & = & e^{-2a|z|}\left(dz^{2}+\tilde{g}_{\mu\nu}(x)dx^{\mu}dx^{\nu}\right)\,,\label{eq:brane}
\end{eqnarray}
where $\tilde{g}_{\mu\nu}$ stands for the metric of a maximally symmetric
spacetime along the four-dimensional brane. Remarkably, a precise
jump in the extrinsic curvature is allowed by the theory in vacuum,
i.e., without the need of an induced stress-energy tensor on the brane.
This effect can also be seen to arise from the fact that the analog
of the Israel junction conditions in this case become quadratic in
the extrinsic curvature, and hence, they admit nontrivial solutions
even in vacuum (see e.g., \cite{Gravanis:2007ei}). As in \cite{Hassaine:2003vq},
metric perturbations along the brane turn out to possess a well-defined
propagator, precisely given by that of a Fierz-Pauli massless graviton,
provided that the induced cosmological constant on the braneworld
is strictly positive. In other words, perturbations of the metric
along the brane reproduce linearized General Relativity around a four-dimensional
de Sitter spacetime with curvature radius given by $a^{-2}$.

It is also worth pointing out that perturbations of the fermionic
fields on the braneworld-like metric \eqref{eq:brane} appear to reproduce
a sort of partially massless version of the Sorokin-Vasiliev doublet
\cite{Sorokin:2008tf} (see also \cite{Deser:2001pe}, \cite{Deser:2001us},
\cite{Skvortsov:2006at}), on de Sitter spacetime \cite{FMT2019WorkInProgess}.

It would also be interesting to study new dimensional reduction schemes
that apply for the class of theories under discussion, as those recently
proposed in \cite{Albornoz:2018uin}.

Formulating hypergravity in presence of cosmological constant would
also be worth to explore. Nonetheless, as in the three-dimensional
case \cite{Chen:2013oxa}, \cite{Zinoviev:2014sza}, \cite{Henneaux:2015ywa},
the introduction of additional bosonic higher spin fields seems to
be inevitable. Indeed, different couplings of higher spin fields to
gravitation on AdS$_{5}$ along these lines have also been proposed
in \cite{Engquist:2007kz}, \cite{Engquist:2008rt}, \cite{Manvelyan:2013oua}.

Besides, the coupling of fermionic fields of half-integer spin with
gravitation in three-dimensional spacetimes is also known to be consistent
\cite{Aragone:1983sz}, and it has been shown that the theory possesses
a suitable reformulation in terms of the hyper-Poincaré algebra with
fermionic generators of spin $s=n+\frac{1}{2}$ \cite{Fuentealba:2015jma}.\footnote{An intriguing link between the hyper-Poincaré algebra with fermionic
generators of half-integer spin and the extended $s$-conformal Galilean
algebra in three dimensions, being purely bosonic, has been recently
pointed out in \cite{Chernyavsky:2019hyp}. Further aspects of the
hyper-Poincaré algebra and its deformations have also recently been
discussed in \cite{Bansal:2018qyz}, \cite{Bansal:2018dgx}.} Lifting these results to higher (odd) dimensions then certainly deserves
consideration. Indeed, for fermionic generators of spin $s=3/2$,
preliminary results suggest that hypergravity theories with a finite
number of bosonic fields can actually be formulated in odd spacetime
dimensions. Interestingly, as in the five-dimensional case, the bosonic
field with the highest spin is also given by $s=3$ in $D=5\text{ mod 4}$
dimensions; while for $D=7\text{ mod }4$, no bosonic higher spin
fields are required ($s\leq2$) \cite{FMT:hyperD}.

\acknowledgments We thank Nicolas Boulanger, Andrea Campoleoni, Marcela
Cárdenas, Hernán González, Daniel Grumiller, Marc Henneaux, Miguel
Pino, Dmitri Sorokin, David Tempo and Stefan Theisen for helpful discussions.
We especially thank Dario Francia and Rakib Rahman for pointing out
the relevance of the ``hook'' fermionic field in the analysis. We
also thank the organizers of the ESI Programme and Workshop \textquotedblleft Higher
Spin and Holography\textquotedblright{} hosted by the Erwin Schrödinger
Institute (ESI), during March of 2019 in Vienna, for the opportunity
of presenting this work. The work of O.F. and J.M. was supported by
the ERC Advanced Grant ``High-Spin-Grav\textquotedbl , by FNRS-Belgium
(convention FRFC PDR T.1025.14 and convention IISN 4.4503.15). This
research has been partially supported by Fondecyt grants Nº 1161311,
1171162, 1181031, 1181496, 3170772. The Centro de Estudios Científicos
(CECs) is funded by the Chilean Government through the Centers of
Excellence Base Financing Program of Conicyt.

\appendix

\providecommand{\href}[2]{#2}\begingroup\raggedright
\bibliographystyle{fullsort} 

\bibliographystyle{unsrt}
\bibliography{review}

\end{document}